\documentclass[a4paper,12pt]{article}

\usepackage{amsmath}
\usepackage{amssymb}
\usepackage{amsthm}
\usepackage[cp1251]{inputenc}
\usepackage[dvips]{graphicx}

\begin{document}

\begin {center}
\noindent{\Large \bf
{Translation invariant theory of polaron (bipolaron) and the problem of quantizing near the classical solution.}}

\medskip
\centerline {Lakhno V.D.}

\medskip
\noindent{\it {Institute of Mathematical Problems of Biology,
Russian Academy of Sciences, Pushchino, Moscow Region, 142290, Russia}}
\end {center}


A physical interpretation of translation-invariant polarons and bipolarons is presented, some results of their existence are discussed.
Consideration is given to the problem of quantization in the vicinity of the classical solution in the quantum field theory.
The lowest variational estimate is obtained for the bipolaron energy
 $E(\eta)$  with
$E(0)=-0,440636\alpha^2$, where $\alpha$  is a constant of electron-phonon coupling, $\eta$  is a parameter of ion binding.

\section{Introduction.}

The quantum field theory is based on the idea that there exist classical solutions in the vicinity of which quantization of fields is realized \cite{lit1}. Such classical solutions can be ordinary plane waves, solitons, kinks, etc. In particular, quantum field theories of a particle interacting with a field proceed from the assumption of the existence of a semiclassical solution (i.e. when a quantum particle moves in a classical field) to which the solution of the quantum field problem must converge in the limit of $\alpha\rightarrow\infty$. In the vicinity of such a solution one can quantize the field and search for the solutions of the quantum field problem thus emerged.

In paper \cite{lit2} we , using the quantum field theory of a strong coupling polaron as an example, demonstrated that, this is not always true and one cannot pass on to the semiclassical description in the case of the limiting transition. This result has a lot of consequences, the most important of which are discussed in this paper.

\section{Interpretation and physical properties of translation-invariant polarons.}

The polaron quantum field translation-invariant theory was constructed in \cite{lit3}. According to this work, the ground state of a translation-invariant polaron is a delocalized state of the electron-phonon system: the probabilities of electron's occurrence at any point of the space are similar. Both the electron density and the amplitudes of phonon modes (renormalized by an interaction with the electron) are delocalized. The concept of a polaron potential well (formed by local phonons  \cite{lit4}) in which the electron is localized, i.e. the self-trapped state is lacking. Accordingly, the induced polarization charge of the translation-invariant polaron is equal to zero.
Polaron's lacking a localized "phonon environment" suggests that its effective mass is not very much different from that of an electron.
The ground state energy of  the translation-invariant polaron is lower than that of  Pekar polaron and is  $E_0=-0,1257520\alpha^2$, \, \cite{lit2} (for Pekar polaron $E_0=-0,10851128\alpha^2$,\,  \cite{lit5}).

Thus, for $P=0$, where $P$ is the total momentum of the polaron there is an energy gap between the translation-invariant polaron state and the Pekar one (i.e. the state with broken translation invariance).  The translation-invariant polaron is a structureless particle (the results of investigations of the Pekar polaron structure are summed up in  \cite{lit4}). According to the translation-invariant polaron theory, the terms "large-radius polaron" (LRP) and "small-radius polaron" (SRP) are relative, since in both cases the electron state is delocalized over the crystal.  The difference between the LRP and SRP in the translation-invariant theory lies in the fact that for the LRP the inequality $k_{char}a< \pi$  is fulfilled, while for the SRP $k_{char}a>\pi$  holds, where $a$ is the lattice constant and $k_{char}$ is a characteristic value of the wave vectors making the main contribution into the polaron energy.
This statement is valid not only for Pekar-Fr\"ohlich polaron, but for the whole class of polarons whose coupling constant is independent of the electron wave vector, such as Holstein polaron for example. For polarons whose coupling constant depends on the electron wave vector, these criteria may not hold (as is the case with  Su-Schreiffer-Heeger model \cite{lit6}).

These properties of translation-invariant polarons determine their physical characteristics which are qualitatively different from those of Pekar polarons. When a crystal has minor local disruptions, the translation-invariant polaron remains delocalized. For example, in an ionic crystal containing lattice vacancies, delocalized polaron states will form $F$-centers only at a certain critical value of the static dielectric constant $\epsilon_{oc}$. For  $\epsilon_o>\epsilon_{oc}$ , a crystal will have delocalized translation-invariant polarons and free vacancies. For  $\epsilon_o=\epsilon_{oc}$ , a transition from the delocalized state to that localized on vacancies (collapse of the wave function) will take place. Such behavior of translation-invariant polarons is qualitatively different from that of Pekar polarons which are localized on the vacancies at any value of  $\epsilon_o $ . This fact accounts for, in particular, why free Pekar polaron does not demonstrate absorption (i.e. structure), since in this case the translation-invariant polaron is realized. Absorption is observed only when a bound Pekar polaron, i.e. $F$-center is formed.
These statements are also supported by a set of recent papers where Holstein polaron is considered \cite{lit7}-\cite{lit9}.

Notice that the physics of only free strong-coupling polarons needs to be changed. The overwhelming majority of results on the physics of strong-coupling polarons have been obtained for bound (on vacancies or lattice disruptions) polaron states of Pekar type and do not require any revision.

Taking account of translation invariance in the case of the polaron leads to a minor change in the value of the ground state and, at first sight, seems irrelevant for physical applications. However, there are two facts which do not let us make this conclusion. First, the estimation of the energy obtained for the translation-invariant polaron, being variational, gives only the upper value of the ground state energy. Second, there are cases when consideration of translation invariance results in much greater changes in the ground state energy. Among them is the case of a bipolaron considered below.

An apt illustration of the aforesaid can be, for example, estimations of energies of a translation-invariant state, a self-localized state and that localized on a static defect that can be close, which however does not imply that the states are identical. Thus, the properties of a polaron localized on a static defect and a light electron-polaron localized on a heavy hole are very much alike, though one of them is translation-invariant, while the other is not (see fig. 6 in \cite{lit10}).

\section{ Translation-invariant bipolarons.}

The quantum-field translation-invariant bipolaron theory was constructed in  \cite{lit11} on the basis of the translation-invariant polaron theory  \cite{lit3}.  According to  \cite{lit11}, the ground state of the translation-invariant bipolaron is delocalized. Here we will show that estimation of the ground state found in  \cite{lit11},  \cite{lit12} with the use of one- and two-parameter probe wave functions can be improved if we use a three-parameter wave function. The consequences of this fact are also discussed in this section.

According to  \cite{lit11}, in the case of the bipolaron, Froehlich Hamiltonian has the form:

\begin{eqnarray}
\label {eq.1}
\hat{H} = -\frac {\hbar ^2}{2M_e}\Delta _{R}-
\frac {\hbar ^2}{2\mu _e}\Delta _{r} +U(|{\vec r}|)\\ \nonumber
+\sum_k{\hbar\omega a^+_k a_k} +\sum_k 2\cos \frac{kr}{2} (V_ka_k e^{ikR} + h.c.),
\end{eqnarray}

\noindent where $\vec{R}$, $\vec{r}$ are the coordinates of the electrons mass-center and electrons relative motion, respectively; $M_e=2m$, $\mu_e=m/2$, $m$ is the electron mass;    $a^+_k$, $a_k$ are operators of the phonon field; $V_k=(e/k)\sqrt{2\pi\hbar\omega /\tilde{\epsilon}V}$, $\tilde{\epsilon}^{-1}=\epsilon^{-1}_{\infty}-\epsilon^{-1}_0$, where $\omega$  is the phonon frequency, $e$ is the electron charge,  $\epsilon_{\infty}$ and $\epsilon_0$  are high-frequency and static dielectric constants, $V$ is the system's volume, $U(r)=e^2/\epsilon_{\infty}|\vec{r}|$. To minimize the total energy $E=\left\langle \Psi_0|\hat{H}|\Psi_0\right\rangle$  we in  \cite{lit11} chose the wave function:

\begin{eqnarray}
\label {eq.2}
\Psi _0 = \Psi (r)\exp \{-i\sum _k\vec k a^+_k a_k \vec R\}
\cdot \exp \{ \sum _k f_k(a_k - a^+_k)\}\Lambda _0,
\end{eqnarray}

\noindent The explicit form of $\Lambda_0=\Lambda_0({f_k},{a_k})$  is given in  \cite{lit11}. Let us choose the probe wave function $\Psi(r)$ and variational parameters $f_k$ in the form:

\begin{eqnarray}
\label {eq.3}
f_k=-N\bar V_k \exp (-k^2/2\mu),\ \psi(r)=(2/\pi l^2)^{3/4}\exp (-r^2/l^2),\\ \nonumber
\end{eqnarray}
$$
{\bar V}_k = 2V_k\langle \Psi | \cos\frac{\vec{k}\vec{r}}{2}| \Psi\rangle,\quad
$$

\noindent where $N$, $\mu$, $l$  are variational parameters. For $N=1$, expression (3) reproduces the results of work  \cite{lit12}, and for $N=1$ and $\mu\rightarrow\infty$, those of work  \cite{lit11}.

Having substituted (2), (3) into the expression for the total energy and then minimized the expression obtained with respect to parameter $N$ we write $E$ as:

\begin{eqnarray}
\label {eq.4}
E(x,y;\eta)=\phi(x,y;\eta)\alpha^2,
\end{eqnarray}

$$
\phi(x,y;\eta)=\frac{6}{x^2}+\frac{20,25}{x^2+16y}
-\frac{16\sqrt{x^2+16y}}{\sqrt{\pi}(x^2+8y)}+\frac{4\sqrt{2/\pi}}{x(1-\eta)},
$$

\noindent where $x$, $y$ are variational parameters: $\alpha=(e^2/\hbar\tilde{\epsilon})\sqrt{m/2\hbar\omega}$,
$x=l\alpha$ , $y= \alpha^2/\mu$,  is the constant of electron-phonon coupling,
$\eta=\epsilon_{\infty}/\epsilon_0$. We assume that  $\hbar=1$,
$\omega=1$, $M_e=1$ (accordingly $\mu_e=1/4$ ). Let us write  $\Phi_{min}$  for the minimum of function
$\phi$ of parameters $x$ and $y$.
Fig.1 shows the dependence of $\Phi_{min}$ on the parameter $\eta$.
Fig.2 demonstrates the dependence of $x_{min}$, $y_{min}$ on the parameter $\eta$.

\begin{figure}
\includegraphics[scale=0.7]{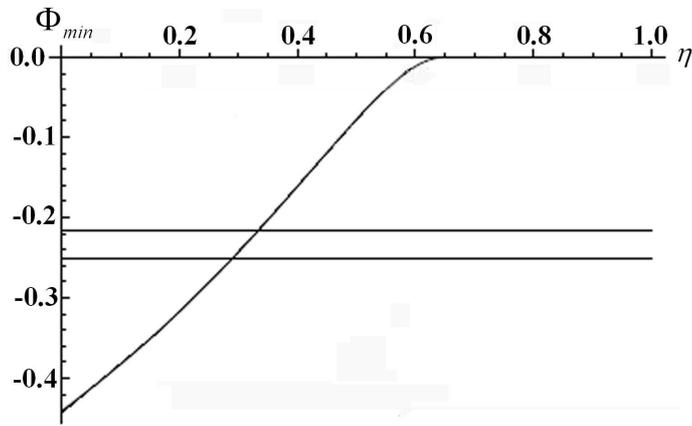}
\caption{Function $\Phi_{min}(\eta)=\min_{x,y}\phi(x,y,\eta)$  and horizontal lines $-0.217$, $-0.2515$;
$\Phi_{min}(0)=-0.440636$.}
\end{figure}

\begin{figure}
\includegraphics[scale=0.7]{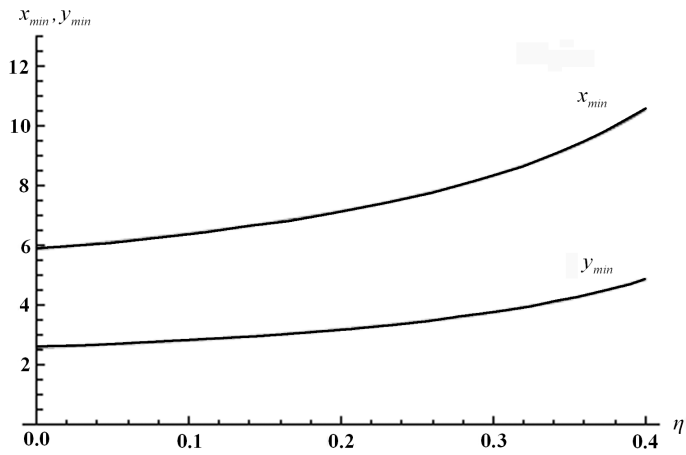}
\caption{Coordinates $x_{min}(\eta)$, $y_{min}(\eta)$ where function has minimum at $0\leq\eta\leq0.4$ }
\end{figure}

Fig.1 suggests that $E_{min}(\eta=0)=-0,440636\alpha^2$  yields the lowest estimation of the bipolaron ground state energy as compared to all those obtained earlier by variational method. Horizontal lines in Fig.1 correspond to the energies: $E_1=-0,217\alpha^2$  and   $E_2=-0,2515\alpha^2$  , where $E_1=2E_{P1}$ , $E_{P1}$ is the Pekar ground state energy  \cite{lit5}; $E_2=2E_{P2}$ is the ground state energy of the translation-invariant polaron  \cite{lit2}. Intersection of these lines with the curve $E_{min}(\eta)$  yields the critical values of the parameters  $\eta=\eta_{C1}=0,3325$ and  $\eta=\eta_{C2}=0,289$. For  $\eta>\eta_{C2}$, the bipolaron decays into two translation-invariant polarons, for $\eta>\eta_{C1}$ , it breaks down into Pekar polarons. The values of minimizing parameters $x_{min}$ and $y_{min}$ for these values of  $\eta$ are
$x_{min}(0)=5,87561$; $y_{min}(0)=2,58537$;
$x_{min}(0,289)=8,16266$; $y_{min}(0,289)=3,68098$;
$x_{min}(0,3325)=8,88739$; $y_{min}(0,3325)=4,03682$;

The critical value of the electron-phonon coupling constant $\alpha$  at which the translation-invariant bipolaron is formed is equal to $\alpha_C=4,54$ , being the lowest estimate obtained by variational method.

Notice, that for the derived ground state of the translation-invariant bipolaron, the virial theorem holds  \cite{lit13}:

\begin{eqnarray}
\label {eq.5}
\frac{\phi_{kin}}{\phi}=-1,\ \frac{\tilde{\phi}_{el}}{\phi}=3,\ \frac{\tilde{\phi} _{int}}{\phi}=4,
\end{eqnarray}

where
$$\phi_{kin}=\frac{20,25}{x^2+16y}+\frac{6}{x^2},\
 \phi_{int}=-\frac{32}{\sqrt{\pi}}\frac{\sqrt{x^2+16y}}{x^2+8y},$$

$$\tilde{\phi}_{int}=\phi_{int} + 2\bar{U},\
\bar{U}=4\frac{\sqrt{2/\pi}}{x(1-\eta)},\
\tilde{\phi}_{el}=\phi_{kin} +\tilde{\phi}_{int}.$$

\noindent The quantities $E_{kin}=\phi_{kin}\alpha^2$,  $E_{int}=\phi_{int}\alpha^2$, $E_{el}=(\phi_{kin}+\phi_{int}+\bar{U})\alpha^2$in (5) have the meaning of the kinetic energy, the energy of the interaction between an electron and a phonon field, and the electron energy, respectively:
$$E_{kin}=\sum_{i=1,2}\left\langle \Psi\left|-\frac{\hbar}{2m}\Delta_i\right|\Psi\right\rangle,\
E_{int}= \sum_{k,i=1,2}\left\langle \Psi\left|\sum_k \left[V_ka_ke^{ikr_i}+h.c.\right]\right|\Psi\right\rangle,$$

$$E_{el}=E_{kin}+E_{int}+\left\langle \Psi\left|U(r)\right| \Psi\right\rangle,\ \bar{U}=\left\langle \Psi\left|U(r)\right|\Psi\right\rangle/\alpha^2.$$

\noindent Relations (5) hold true with an accuracy of the sixth decimal place.  The great binding energy of the translation-invariant bipolaron has important physical consequences. In particular, when a crystal has minor local disruptions, the translation-invariant bipolaron will be delocalized. Thus, in an ionic crystal with lattice vacancies, formation of  $F'$ -centers by delocalized bipolarons will take place only at a certain critical value of the static dielectric constant $\epsilon_{oc}$.  For  $\epsilon_o>\epsilon_{oc}$ , the crystal will contain delocalized translation-invariant bipolarons and free vacancies.

In the case of  $\epsilon_o=\epsilon_{oc}$ translation-invariant bipolarons will pass on from the delocalized state to that localized on vacancies, i.e. to  $F'$- center.  Such behavior of translation-invariant bipolarons is qualitatively different from the behavior of Pekar polarons with spontaneously broken symmetry  \cite{lit13}, which are localized on the vacancies at any values of  $\epsilon_o$.

In the case of $P=0$, where $P$ is the total momentum of the bipolaron, the translation-invariant bipolarons, being delocalized, will be separated from those with broken translation invariance  by an energy gap. In  \cite{lit12} a suggestion was made that translation-invariant polaron and bipolaron states are superconducting.
As with a translation-invariant polaron, in the case of a translation-invariant bipolaron there is no reason to believe that its mass differs greatly from the double mass of paired electrons.
As is known, interpretation of the high-temperature superconductivity relying on the bipolaron mechanism of Bose-condensation runs into a problem associated with a great mass of bipolarons and, consequently, low temperature of Bose-condensation.
The possibility of a smallness of translation-invariant bipolaron's mass resolves this problem. It should be stressed that
the above-mentioned properties of translation-invariant bipolarons impart them superconducting properties even in the absence of Bose-condensation.

\section{{%
Discussion about completeness of Tulub's theory.}}

In \cite{lit14}, \cite{lit15} a question was raised as to whether Tulub's theory \cite{lit3} is complete. Arguments of \cite{lit14}, \cite{lit15} are based on the work by Porsch and R\"oseler \cite{lit16} which reproduces the results of Tulub's theory. However, in the last section of their paper Porsch and R\"oseler investigate what will happen if the infinite integration limit in Tulub's theory change for a finite limit and then pass on to an infinite one. Surprisingly, it was found that in this case in-parallel with cutting of integration to phonon wave vectors in the functional of the polaron total energy one should augment the latter with the addition $\delta E^{PR}$ which will not disappear if the upper limit tends to infinity \cite{lit15}, \cite{lit17}. Relying on this result the authors of \cite{lit14}, \cite{lit15} concluded that Tulub did not take this addition into account and therefore his theory is incomplete.
To resolve the paradox with the contribution of "plasma frequency" into the polaron energy which leads
to introducing $\delta E^{PR}$ addition \cite{lit16}, let us consider the function $\Pi _q(s)$ determined by formula (37c)
in paper \cite{lit16}. Its zeros contribute into the polaron recoil energy and, according to \cite{lit16},
are found from the solution of the equation:

\begin{eqnarray}
\label {eq.6}
1=\lambda\frac{2}{3m}\sum_q\frac{q^2f^2_q\hbar\omega_q}{s-\omega^2_q},
\end{eqnarray}
where the notations are the same as that used in \cite{lit16}.

If a cutoff is absent in the sum on the right-hand side of Eq (6), then the solution of Eq (6)
yields a spectrum of $s$ values determined by frequencies  $\Omega _{q_i}$ lying between neighboring values of $\omega _{q_i}$ and
  $\omega _{q_{i+1}}$ for all the wave vectors  $q_i$. These frequencies determine the value of the polaron recoil energy:

\begin{eqnarray}
\label {eq.7}
\Delta E=\frac{\hbar}{2}\sum_{q_i}(\Omega_{q_i}-\omega_{q_i})
\end{eqnarray}

\noindent Let us see what happens with the contribution of frequencies $\Omega _{q_i}$ into $\Delta E$ in the
region of the wave vectors
$q$ where  $f_q$ vanishes but nowhere becomes exactly
zero. From (6) it follows that as $f_q\rightarrow 0$, solutions of equation
(6)  $\Omega _{q_i}$ will tend to $\omega _{q_i}$: $\Omega _{q_i}\rightarrow\omega _{q_i}$. Accordingly, the contribution of the
wave vectors region into $\Delta E$, where $f_q\rightarrow 0$,
will also tend to zero.

In particular, if we introduce a certain $q^0$ such that in the region $q>q^0$ the values of $f_q$ are small,
we will express $\Delta E$ in the form:

\begin{eqnarray}
\label {eq.8}
\Delta E=\frac{\hbar}{2}\sum_{q_i\leq q^0}(\Omega_{q_i}-\omega_{q_i})
\end{eqnarray}
which does not contain any additional terms. To draw a parallel with Tulub's approach,
there we could put the upper limit $q^0$, but no additional terms would appear.

For example, if in an attempt to investigate the minimum of Tulub's functional we choose
 the probe function  $f_q$ not containing a cutoff in the form:

\begin{eqnarray}
\label {eq.9}
f_q=-V_q \exp (-q^2/2a^2(q)),\\ \nonumber
a(q)=\frac{a}{2}\left[1+th\left(\frac{q_b-q}{a}\right)\right]
\end{eqnarray}
where $a$ is a parameter of Tulub's probe function \cite{lit3} and  $q_b$ satisfies the condition
$a<<q_b<<q_{oc}$, $q_{0c}$ - is the value of the wave vector for which Tulub's integral $q(1/\lambda)$ has a maximum \cite{lit3}, \cite{lit17}
 then with the use of (9) in the limit   $\alpha\rightarrow\infty$,
 Tulub's integral $q(1/\lambda)$
 will be written as:

\begin{eqnarray}
\label {eq.10}
q(1/\lambda)\cong 5,75 + 6(a/q_b)^3exp(-q^2_b/a^2)
\end{eqnarray}
The second term in the right-hand side of the expression for $q(1/\lambda)$ vanishes as $(q_b/a)\rightarrow\infty$ and we get,
as we might expect, Tulub's result: $q(1/\lambda)\cong5,75$.

Equation (6), however, has a peculiarity. Even in the case of a continuous spectrum, for $f_q=0$,
if $q>q^0$, it has an isolated solution $\Omega _{q^0}$ separated from the maximum frequency $\omega _{q^0}$
 by a finite value.
This isolated solution leads to an additional contribution $\delta E^{PR}$ into  $\Delta E$:

\begin{eqnarray}
\label {eq.11}
\Delta E = \frac{1}{2}\sum_{q_i<q^0}(\Omega _{q_i}-\omega _{q_i})+\delta E^{PR},\\ \nonumber
\delta E^{PR} = \frac{3}{2}(\Omega _{q^0}-\omega _{q^0})
\end{eqnarray}
where $\Omega _{q^0}$ has the meaning of "plasma frequency". Hence here a continuous transition from the case
of $f_q\rightarrow 0$ for  $q>q^0$ to the case of $f_q=0$ for  $q>q^0$ is absent.
As is shown by the direct calculation \cite{lit17}
of the contribution of the terms with "plasma frequency" $\delta E^{PR}$  into (11), even for $q^0\rightarrow\infty$,
Porsch  and R\"oseler  theory does not transform itself into Tulub's theory.

In Tulub's theory we choose such $f_q$ which lead to the minimum of the functional of the polaron
total energy. In particular, the choice of the probe function in the form (9) provides the absence
of a contribution from the "plasma frequency" into the total energy and in actual calculations one
can choose a cutoff quantity $f_q$  without introducing any additional terms in Tulub's functional \cite{lit18}.

\section{Conclusive remarks}
In papers \cite{lit14}, \cite{lit15} some critical remarks were made concerning Tulub's theory \cite{lit3} and the author's works based on it \cite{lit11}, \cite{lit12}. Inadequacy of this criticism was discussed in \cite{lit17}, \cite{lit18} and herein.
Presently Tulub's theory and the quantitative results \cite{lit11}, \cite{lit12} obtained on its basis are undoubted.

An important role in checking the validity of the theory belongs to virial relations. In paper \cite{lit2} it was demonstrated that in Tulub's theory these relations are strictly fulfilled for the case of a polaron. Here we have shown that they are strictly fulfilled for a bipolaron too. Notice, that the expressions derived in \cite{lit14}, \cite{lit15} for the total energy of a polaron and bipolaron do not satisfy virial relations.

The quantum-field theory constructed in  \cite{lit3} is nonperturbative. There, the strong binding limit is achieved by choosing an appropriate wave function. If the wave function is chosen to be more complicated, the theory can reproduce the modes of intermediate and weak coupling \cite{lit19}. Presently, it is believed that one of the most effective techniques for calculating polarons and bipolarons in this range of interaction is the method of path integrals. This approach
not being properly modified  is not translation-invariant since
 there the main contribution into the energies is made by classical solutions (i.e.  extrema points of the exponent of classical action involved in the path integral).
Though, in view of translation invariance, such solutions are not isolated stationary points, but belong to a continuous family of classical solutions obtained as a result of the action of a translation operator on the initial classical solution. Accordingly, without a proper modification, the stationary phase approximation is nonapplicable in a translation invariant system.

In the quantum field theory some approaches based on the introduction of collective coordinates in the functional integral have been developed to restore the translation invariance  \cite{lit1}. However, up to now they have not been used in the polaron theory. For this reason it is no wonder that  the method of path integrals, as used in the polaron theory, yields a result which coincides with that obtained in the semiclassical theory of the strong-coupling polaron  \cite{lit20}.

In conclusion, the author expresses gratitude to A.V. Tulub and N. I. Kashirina for discussion of various aspects of the problems considered in the paper.

The work was done with the support from the RFBR, projects N 11-07-12054, 10-07-00112.


\end{document}